\batchmode
\makeatletter
\def\input@path{{"E:/IOP Research/manuscripts/Nb3Cl8/PRL Final Submission/"}}
\makeatother
\documentclass[twocolumn,prl,showpacs,superscriptaddress,longbibliography]{revtex4-2}
\usepackage{textcomp}
\usepackage[latin9]{inputenc}
\usepackage{color}
\usepackage{amsmath}
\usepackage{graphicx}
\usepackage[pdftex,pdfusetitle,
 bookmarks=true,bookmarksnumbered=false,bookmarksopen=false,
 breaklinks=false,pdfborder={0 0 0},pdfborderstyle={},backref=false,colorlinks=true]
 {hyperref}

\makeatletter

\newcommand{\lyxmathsym}[1]{\ifmmode\begingroup\def\b@ld{bold}
  \text{\ifx\math@version\b@ld\bfseries\fi#1}\endgroup\else#1\fi}

\makeatother

\begin{document}
\title{Observation of Dimensionality-dependent Exciton Dispersion in a Single-Band
Mott Insulator}
\affiliation{Beijing National Laboratory for Condensed Matter Physics and Institute
of Physics, Chinese Academy of Sciences, Beijing 100190, China}
\affiliation{School of Physics, Zhejiang University, Hangzhou, 310058, China}
\affiliation{School of Physics and Beijing Key Laboratory of Opto-electronic Functional
Materials \& Micro-nano Devices, Renmin University of China, Beijing
100872, China}
\affiliation{School of Physical Sciences, University of Chinese Academy of Sciences,
Beijing 100049, China}
\affiliation{State Key Laboratory of Silicon and Advanced Semiconductor Materials,
Zhejiang University, Hangzhou, 310027, China}
\affiliation{Hefei National Laboratory, Hefei 230088, China}
\affiliation{Key Laboratory of Quantum State Construction and Manipulation (Ministry
of Education), Renmin University of China, Beijing, 100872, China}
\affiliation{International Center for Quantum Materials, and Electron Microscopy
Laboratory, School of Physics, Peking University, Beijing 100871,
China}
\affiliation{Department of Physics, Fuzhou University, Fuzhou 350108, Fujian, China}
\author{Zhibin Su}
\affiliation{Beijing National Laboratory for Condensed Matter Physics and Institute
of Physics, Chinese Academy of Sciences, Beijing 100190, China}
\affiliation{School of Physical Sciences, University of Chinese Academy of Sciences,
Beijing 100049, China}
\author{Junjian Mi}
\affiliation{School of Physics, Zhejiang University, Hangzhou, 310058, China}
\affiliation{State Key Laboratory of Silicon and Advanced Semiconductor Materials,
Zhejiang University, Hangzhou, 310027, China}
\author{Shaohua Yan}
\affiliation{School of Physics and Beijing Key Laboratory of Opto-electronic Functional
Materials \& Micro-nano Devices, Renmin University of China, Beijing
100872, China}
\affiliation{Key Laboratory of Quantum State Construction and Manipulation (Ministry
of Education), Renmin University of China, Beijing, 100872, China}
\author{Jiade Li}
\affiliation{Beijing National Laboratory for Condensed Matter Physics and Institute
of Physics, Chinese Academy of Sciences, Beijing 100190, China}
\affiliation{International Center for Quantum Materials, and Electron Microscopy
Laboratory, School of Physics, Peking University, Beijing 100871,
China}
\author{Siwei Xue}
\affiliation{Beijing National Laboratory for Condensed Matter Physics and Institute
of Physics, Chinese Academy of Sciences, Beijing 100190, China}
\affiliation{Department of Physics, Fuzhou University, Fuzhou 350108, Fujian, China}
\author{Zhiyu Tao}
\affiliation{Beijing National Laboratory for Condensed Matter Physics and Institute
of Physics, Chinese Academy of Sciences, Beijing 100190, China}
\affiliation{School of Physical Sciences, University of Chinese Academy of Sciences,
Beijing 100049, China}
\author{Enling Wang}
\affiliation{Beijing National Laboratory for Condensed Matter Physics and Institute
of Physics, Chinese Academy of Sciences, Beijing 100190, China}
\affiliation{School of Physical Sciences, University of Chinese Academy of Sciences,
Beijing 100049, China}
\author{Xiongfei Shi}
\affiliation{Beijing National Laboratory for Condensed Matter Physics and Institute
of Physics, Chinese Academy of Sciences, Beijing 100190, China}
\affiliation{School of Physical Sciences, University of Chinese Academy of Sciences,
Beijing 100049, China}
\author{Hechang Lei}
\email{hlei@ruc.edu.cn}

\affiliation{School of Physics and Beijing Key Laboratory of Opto-electronic Functional
Materials \& Micro-nano Devices, Renmin University of China, Beijing
100872, China}
\affiliation{Key Laboratory of Quantum State Construction and Manipulation (Ministry
of Education), Renmin University of China, Beijing, 100872, China}
\author{Zhuan Xu}
\email{zhuan@zju.edu.cn}

\affiliation{School of Physics, Zhejiang University, Hangzhou, 310058, China}
\affiliation{State Key Laboratory of Silicon and Advanced Semiconductor Materials,
Zhejiang University, Hangzhou, 310027, China}
\affiliation{Hefei National Laboratory, Hefei 230088, China}
\author{Jiandong Guo}
\email{jdguo@iphy.ac.cn}

\affiliation{Beijing National Laboratory for Condensed Matter Physics and Institute
of Physics, Chinese Academy of Sciences, Beijing 100190, China}
\affiliation{School of Physical Sciences, University of Chinese Academy of Sciences,
Beijing 100049, China}
\author{Xuetao Zhu}
\email{xtzhu@iphy.ac.cn}

\affiliation{Beijing National Laboratory for Condensed Matter Physics and Institute
of Physics, Chinese Academy of Sciences, Beijing 100190, China}
\affiliation{School of Physical Sciences, University of Chinese Academy of Sciences,
Beijing 100049, China}
\begin{abstract}
Excitonic band structure is critical for investigating exciton dynamics.
Theoretically, quantum effects from exchange scattering between electron-hole
pairs significantly modulate exciton dispersion. Here, we report the
direct observation of dimensionality-dependent exciton dispersion
in a single-band Mott insulator Nb\textsubscript{3}Cl\textsubscript{8}
through High-Resolution Electron Energy Loss Spectroscopy. In the
high-temperature phase, the exciton in Nb\textsubscript{3}Cl\textsubscript{8}
hosts an exceptionally large binding energy, and exhibits clear quasi-two-dimensional
massless linear dispersion. In contrast, in the low-temperature phase,
the exciton splits into two bands, both displaying three-dimensional
parabolic dispersion. These dramatic changes in the exciton dispersion
stem from the dimensional mutation driven by a substantial enhancement
of interlayer coupling across the phase transition. This study provides
a clear and typical example of how exciton behavior evolves with dimensionality.
\end{abstract}
\maketitle
Excitons, which are electron-hole pairs bound by Coulomb interaction,
play a crucial role in the optoelectronic properties of semiconductors
\citep{RN586,RN588,RN587}. Conventionally, research on excitonic
properties has primarily focused on bright excitons, which can be
directly excited by light \citep{RN588,RN216}. In recent years, there
has been growing interest in dark excitons \citep{RN233,RN612,RN216,RN624,RN623},
especially the momentum-forbidden dark excitons \citep{RN223,RN233,RN235},
which greatly influence the exciton dynamics. The bright excitons
and the momentum-forbidden dark excitons can be considered as distinct
parts of excitonic band structures. The excitonic band structure has
significant impacts on the performance of semiconductors, influencing
key properties such as radiative lifetime \citep{RN592,RN617,RN618,RN593},
quantum yield \citep{RN236,RN618}, and exciton diffusion coefficient
\citep{RN221,RN595,RN596}. Therefore, understanding the excitonic
energy band structure is essential for gaining deep insights into
exciton dynamics.

\begin{figure}
\includegraphics[width=0.45\textwidth]{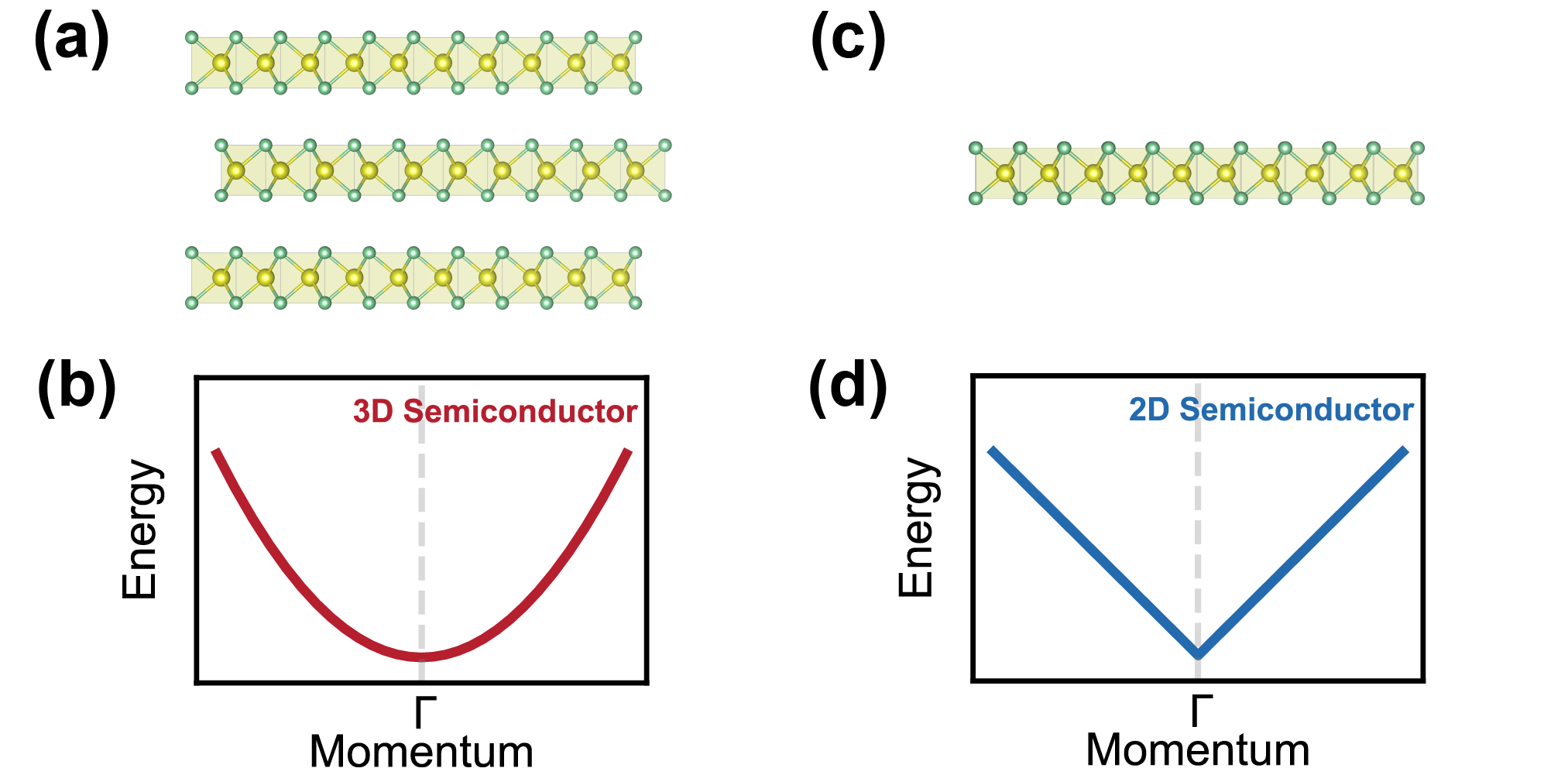}

\caption{\label{fig:Schematic of Dimensionality effect}\textbf{Illustration
of the dimensional effects of excitonic band dispersions.} Schematic
diagrams of (a) the crystal structures and (b) the corresponding excitonic
band structure near the BZ center for typical 3D semiconductors. (c)
Crystal structures and (d) the corresponding excitonic band structure
for typical 2D semiconductors.}
\end{figure}

Microscopically, due to the quantum effects arising from exchange
scattering between electron-hole pairs, the dispersion of an excitonic
energy band is strongly modulated by the dimensionality of the system.
Macroscopically, this dependence arises because the dimensionality
influences the Coulomb screening strength through the system's dielectric
environment. In three-dimensional (3D) semiconductors {[}Fig. \ref{fig:Schematic of Dimensionality effect}(a){]},
the strong Coulomb screening inherent to the 3D space results in Wannier
excitons displaying a parabolic dispersion. {[}Fig. \ref{fig:Schematic of Dimensionality effect}(b){]}
\citep{RN221,RN587}. In contrast, when the system's dimensionality
is reduced to two-dimensional (2D) {[}Fig. \ref{fig:Schematic of Dimensionality effect}(c){]},
the Coulomb screening is confined to the 2D plane, and thus the excitons
exhibit distinctive massless linear dispersions near the Brillouin
zone (BZ) center {[}Fig. \ref{fig:Schematic of Dimensionality effect}(d){]}
------a phenomenon predicted in archetypical 2D materials such as
h-BN, monolayer MoS2, and black phosphorus \citep{RN567,RN238,RN221,RN596,RN576}.
In the 2D limit, the loss of massive character not only suppresses
the excitonic density of states at the BZ center but also endows the
excitons with high, constant group and phase velocities. Consequently,
these unique features enable 2D massless excitons to exhibit shorter
radiative lifetimes \citep{RN221,RN617}, enhanced diffusion rates
\citep{RN221,RN595}, weaker exciton-phonon scattering \citep{RN595},
and the potential for unconventional exciton superfluity \citep{RN597}.

Recently, van der Waals (vdW) materials have emerged as a pivotal
platform for investigating excitonic dimensional effects and many-body
interactions, owing to their reduced dielectric screening and interlayer
connections through weak vdW forces \citep{RN228,RN605,RN216}. Using
prototypical vdW materials, experiments have confirmed the parabolic
dispersions of Wannier excitons in bulk systems \citep{RN620,RN214,RN607}.
Meanwhile, twisted light has been used to indirectly verify the existence
of massless exciton dispersion in monolayer MoS\textsubscript{2}
\citep{RN598}. However, investigations of monolayer WSe\textsubscript{2}
using momentum-resolved electron energy loss spectroscopy in a transmission
electron microscope (TEM-EELS) have reported a parabolic dispersion
\citep{RN217}, which conflicts with theoretical predictions. This
apparent discrepancy can be attributed to two concurrent factors:
the limited momentum and energy resolution of the technique \citep{RN648},
and the confinement of the linear dispersion to an extremely small
momentum region around the BZ center. Together, these limitations
prevented the experimental resolution of the linear dispersion feature.
Consequently, direct and robust experimental verification of massless
exciton dispersion in 2D materials remains lacking, let alone the
observation of dimensional effects on exciton dispersion within the
same material, which presents an even greater challenge.

Here, we employ High-Resolution Electron Energy Loss Spectroscopy
(HREELS) with 2D momentum-energy mapping capabilities \citep{RN604}
to investigate exciton dispersions across the phase transition in
Nb\textsubscript{3}Cl\textsubscript{8} -- a textbook single-band
Mott insulator recently proposed theoretically \citep{RN534,RN212,RN213}
and validated experimentally \citep{RN522}. Across the phase transition,
we clearly observed the splitting of excitons. Notably, the exciton
in the high-temperature phase ($\alpha$ phase) exhibits unambiguous
quasi-2D massless dispersion, while in the low-temperature phase ($\beta$
phase), the excitons display 3D parabolic dispersion. This drastic
change in exciton dispersion arises from the change in system dimensionality,
induced by variations in interlayer coupling strength across the phase
transition.

\begin{figure}
\includegraphics[width=0.45\textwidth]{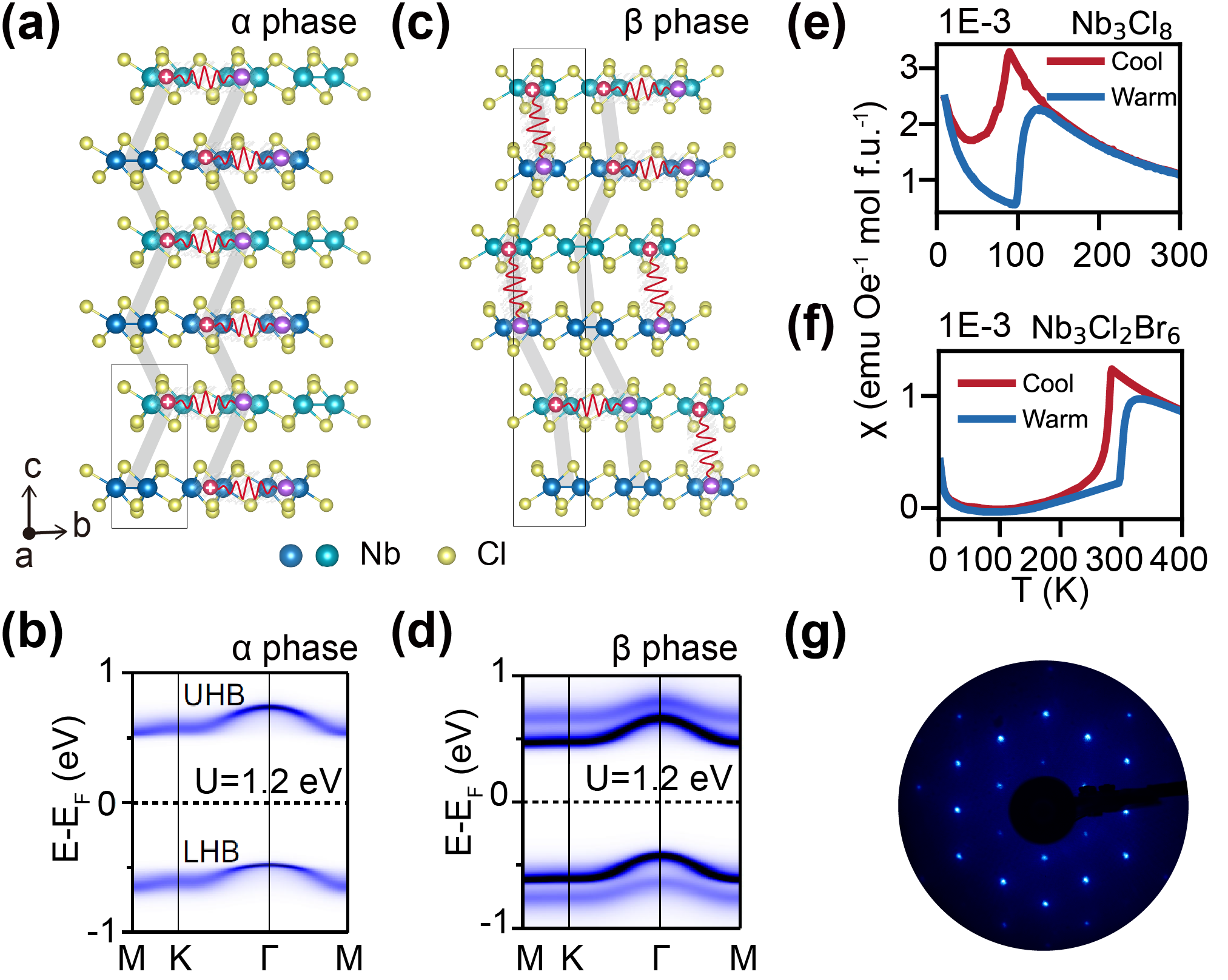}

\caption{\label{fig:Schematic of crystal structure}\textbf{Structural and
electronic properties of Nb\protect\textsubscript{3}Cl\protect\textsubscript{8}.}
(a) Side view of the crystal structure and exciton distribution, and
(b) the corresponding electronic energy band in the high-temperature
phase ($\alpha$ phase) of Nb\protect\textsubscript{3}Cl\protect\textsubscript{8}.
(c) Crystal structure and exciton distribution, and (d) the corresponding
electronic energy band in the low-temperature phase ($\beta$ phase).
(e), (f) Magnetic susceptibility as a function of temperature for
Nb\protect\textsubscript{3}Cl\protect\textsubscript{8} and Nb\protect\textsubscript{3}Cl\protect\textsubscript{2}Br\protect\textsubscript{6}.
(g) LEED pattern on the (001) surface of Nb\protect\textsubscript{3}Cl\protect\textsubscript{8},
obtained at room temperature with incident beam energy of $100$ eV.
Notice that the electronic band structure in (b) and (d) are reproduced
from ref. \citep{RN522}. Copyright 2025 by the American Physical
Society.}
\end{figure}

\emph{ Background of $Nb_{3}Cl_{8}$.} At room temperature, Nb\textsubscript{3}Cl\textsubscript{8}
is in its $\alpha$ phase, where each unit cell consists of two monolayers
stacked along the c-axis via vdW forces {[}Fig. \ref{fig:Schematic of crystal structure}(a){]}
\citep{RN209,RN205}. Within each monolayer, Nb ions spontaneously
form Nb\textsubscript{3} trimers, resulting in a breathing kagome
lattice \citep{RN209,RN584,RN205}, while weak $pd$ hybridization
causes electrons to become strongly localized within the trimers \citep{RN522}.
Concurrently, strong electron correlation leads to the opening of
a Mott gap, which in turn provides the pathway for exciton formation,
as an inter-gap transition can also generate an electron-hole pair
across the Mott gap. The Nb\textsubscript{3} trimers in adjacent
layers are staggered, giving rise to negligible interlayer coupling.
These features enable the electronic properties to be effectively
described by a monolayer \citep{RN522,RN198}, with excitons primarily
confined to the single layer {[}Fig. \ref{fig:Schematic of crystal structure}(a){]}.
Thus, in the $\alpha$ phase, Nb\textsubscript{3}Cl\textsubscript{8}
serves as an ideal quasi-2D single-band Mott insulator, with the valence
and conduction bands arising from the Lower Hubbard Band (LHB) and
Upper Hubbard Band (UHB), both driven by Nb $d$-orbitals {[}Fig.
\ref{fig:Schematic of crystal structure}(b){]}.

As the temperature decreases, Nb\textsubscript{3}Cl\textsubscript{8}
undergoes a structural phase transition near 100 K {[}Fig.\ref{fig:Schematic of crystal structure}
(e){]}, transforming from the paramagnetic $\alpha$ phase to the
non-magnetic $\beta$ phase \citep{RN209,RN210,RN205}. Although the
exact structure of the $\beta$ phase remains under debate, it is
clear that interlayer sliding is a key factor. Interlayer sliding
aligns the Nb\textsubscript{3} trimers in adjacent layers directly
above one another within each bilayer, significantly enhancing the
interlayer coupling {[}Fig. \ref{fig:Schematic of crystal structure}(c){]}.
Therefore, the electronic properties of the $\beta$ phase are best
described as quasi-3D bilayer Nb\textsubscript{3}Cl\textsubscript{8},
with the significantly enhanced interlayer coupling causing bonding-antibonding
splitting in the LHB and UHB {[}Fig. \ref{fig:Schematic of crystal structure}(d){]}
\citep{RN522,RN534}. In this case, the excitons are chiefly located
between the two layers {[}Fig. \ref{fig:Schematic of crystal structure}(c){]}.
Thus, with its uniquely tunable crystal structure, Nb\textsubscript{3}Cl\textsubscript{8}
serves as a superior model system over conventional semiconductors
for probing exciton dimensional effects.

\begin{figure}
\includegraphics[width=0.45\textwidth]{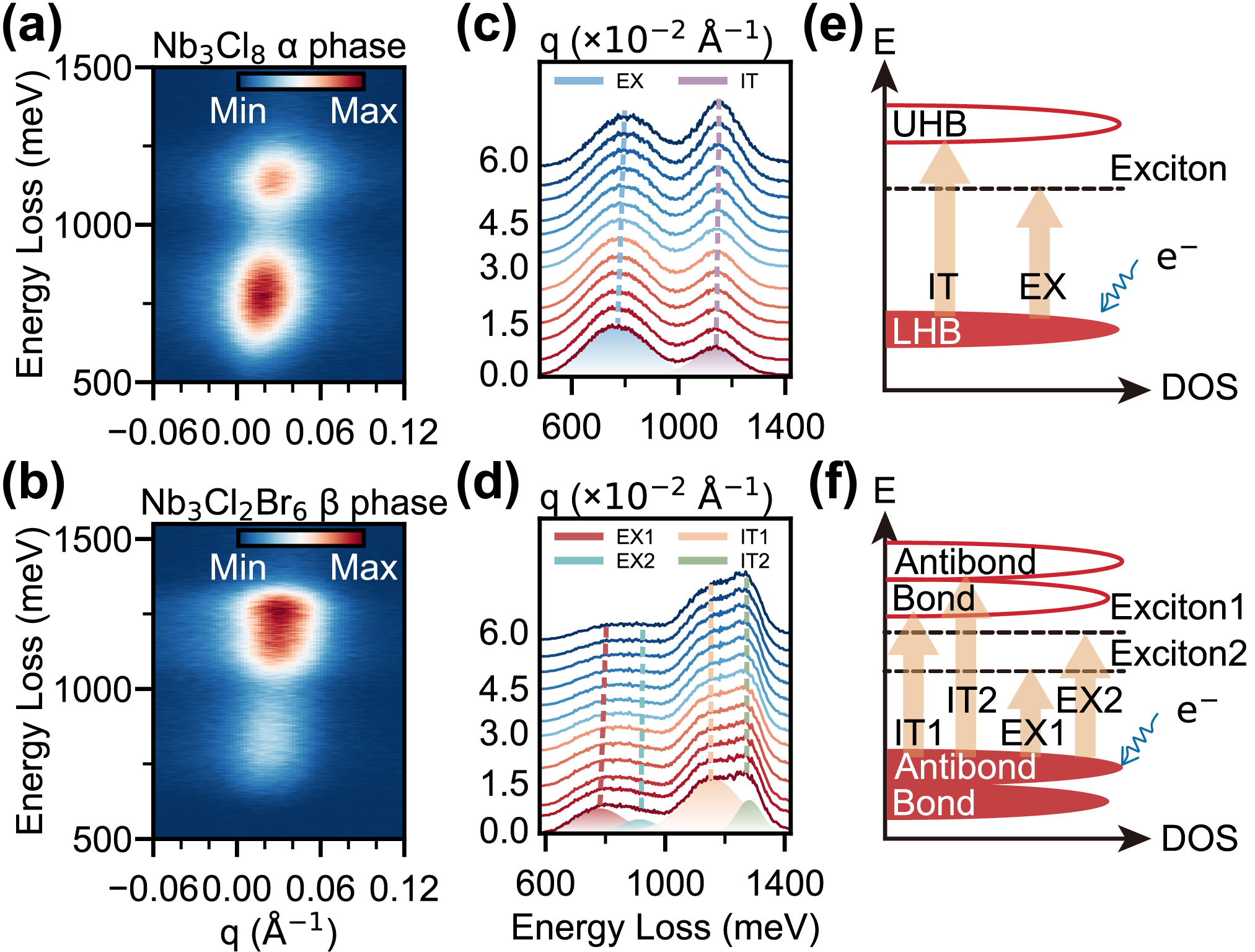}

\caption{\label{fig:Exciton Splitting}\textbf{Exciton splitting from the HREELS
measurements.} (a), (b) Representative 2D momentum-energy mappings
of HREELS along the $\overline{{\rm \Gamma}}\overline{{\rm K}}$ direction
for Nb\protect\textsubscript{3}Cl\protect\textsubscript{8} at 300
K ($\alpha$ phase) and Nb\protect\textsubscript{3}Cl\protect\textsubscript{2}Br\protect\textsubscript{6}
at 106 K ($\beta$ phase). (c), (d) Stacks of momentum-dependent EDCs
along $\overline{{\rm \Gamma}}\overline{{\rm K}}$ direction, with
fitting results for the $\alpha$ phase and $\beta$ phase. The corresponding
momentum values are indicated on the left. For clarity, the EDCs were
multiplied by a scaling factor and shifted vertically. (e), (f) Schematic
representation of the density of state and corresponding electron
excitations for the $\alpha$ phase and $\beta$ phase. Inter-band
transitions and excitons are denoted by IT and EX, respectively.}
\end{figure}

\emph{HREELS measurements.} As Nb\textsubscript{3}Cl\textsubscript{8}
is a semiconductor with high electrical resistance as the temperature
decreases, significant charging effects occur when performing the
HREELS measurements below 250 K, which hinder the observations of
the $\beta$ phase. To diminish the issue of sample charging, we employed
the Nb\textsubscript{3}Cl\textsubscript{2}Br\textsubscript{6} sample.
The substitution of Br with Cl elevates the phase transition temperature
to near room temperature {[}Fig. \ref{fig:Schematic of crystal structure}(f){]}.
Simultaneously, this substitution narrows the band gap, which in turn
effectively suppresses the charging effect \citep{SM}. The measured
HREELS results of the $\alpha$ phase in Nb\textsubscript{3}Cl\textsubscript{2}Br\textsubscript{6}
show features identical to those in Nb\textsubscript{3}Cl\textsubscript{8},
with only minor energy shifts due to doping {[}details described in
the Supplementary Materials (SM) \citep{SM}{]}. Here in the main
text, we present the $\alpha$ phase data from Nb\textsubscript{3}Cl\textsubscript{8}
samples, while the $\beta$ phase data from Nb\textsubscript{3}Cl\textsubscript{2}Br\textsubscript{6}
samples. Nb\textsubscript{3}Cl\textsubscript{8} and Nb\textsubscript{3}Cl\textsubscript{2}Br\textsubscript{6}
single crystals were cleaved \textit{in-situ} under ultra-high vacuum,
and their surfaces were characterized by low energy electron diffraction
(LEED), which shows sharp bright patterns indicating good surface
quality {[}Fig. \ref{fig:Schematic of crystal structure}(g){]}. All
the HREELS measurements were performed with an incident electron energy
of 110 eV and an incident angle of $60^{\circ}$.

\emph{Exciton Splitting.} Along the $\overline{{\rm \Gamma}}\overline{{\rm K}}$
direction, the 2D energy-momentum mapping of HREELS for the $\alpha$
phase {[}Fig. \ref{fig:Exciton Splitting}(a){]} and the $\beta$
phase {[}Fig. \ref{fig:Exciton Splitting}(b){]} reveals several significant
loss signals below 1.5 eV. In the $\alpha$ phase, two well-defined
loss peaks are presented, while in the $\beta$ phase, the shape and
intensity of these loss signals change dramatically. To better illustrate
these features, we extract the momentum-dependent energy distribution
curves (EDCs) for both the $\alpha$ phase {[}Fig. \ref{fig:Exciton Splitting}(c){]}
and the $\beta$ phase {[}Fig. \ref{fig:Exciton Splitting}(d){]},
fitting the EDCs at the $\overline{\Gamma}$ point. Below 1.5 eV,
the fitting results for the $\alpha$ phase show two peaks: a 1.14
eV peak corresponding to inter-band transitions (IT), and a 0.77 eV
peak is attributed to the excitation of an exciton (EX), with the
identification detailed in the SM \citep{SM}. These results indicate
an optical band gap $E_{O}=0.77\ {\rm eV}$ and an electronic band
gap $E_{G}=1.14\ {\rm eV}$, corresponding to excitations from the
LHB to the excitonic band and to the UHB, respectively {[}Fig. \ref{fig:Exciton Splitting}(e){]}.
Then, the exciton in the $\alpha$ phase exhibits a substantial binding
energy of approximately $E_{B}=0.37\ {\rm eV}$, with $E_{B}/E_{G}\sim1/3$,
in accordance with the scaling universality for monolayer 2D semiconductors
\citep{RN578}.

HREELS spectra show four peaks in the $\beta$ phase, suggesting noticeable
splitting in both of the inter-band transition and exciton peaks {[}Fig.
\ref{fig:Exciton Splitting}(d){]} with the discussion detailed in
the SM\citep{SM}. These splittings arise from the bonding-antibonding
splitting in the LHB and UHB in the $\beta$ phase. As shown in Fig.
\ref{fig:Exciton Splitting}(f), the splitting of the UHB generates
two corresponding excitonic bands. Subsequently, hopping from the
LHB antibonding band to the unoccupied bands will generate four different
possible excitations: IT1 and IT2, corresponding to transitions to
the UHB bonding and antibonding bands, and EX1 and EX2, corresponding
to excitations to the two excitonic bands {[}Fig. \ref{fig:Exciton Splitting}(f){]}.

\begin{figure*}
\includegraphics[width=0.95\textwidth]{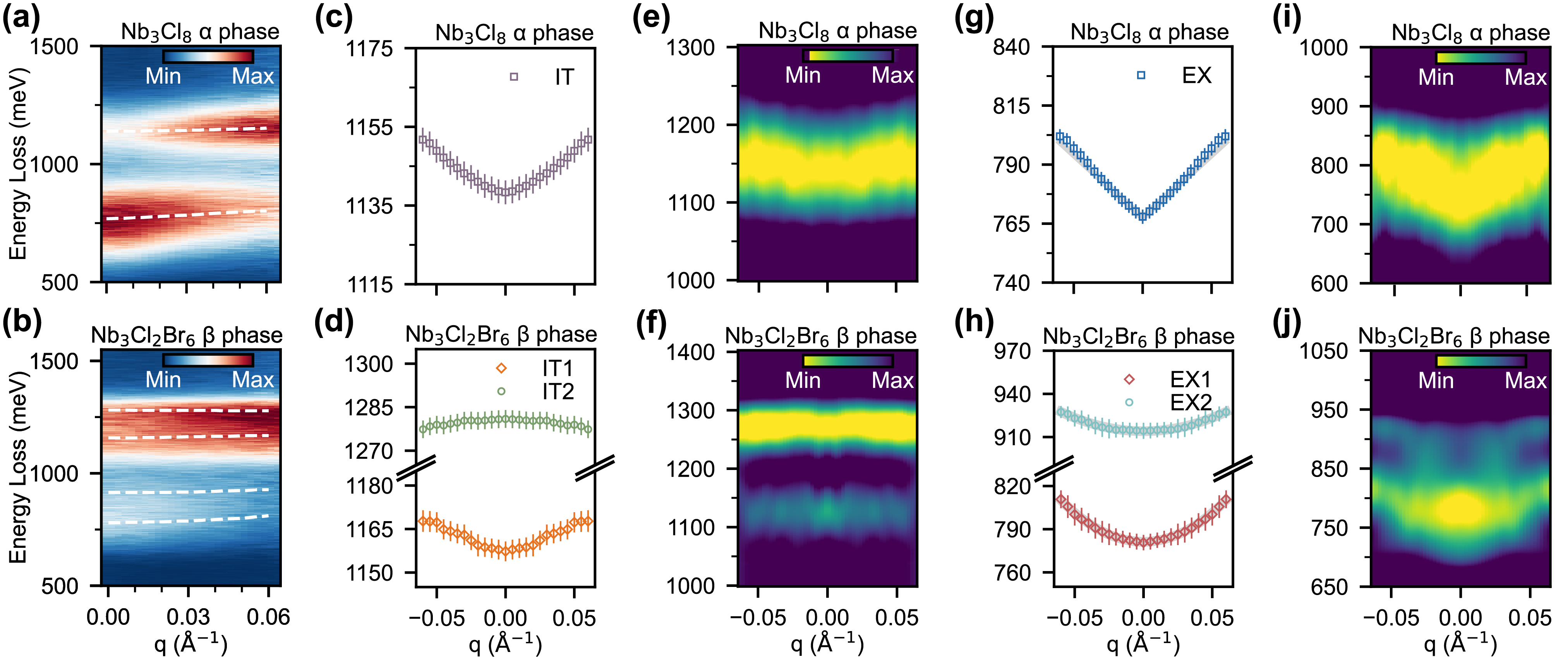}

\caption{\label{fig:Non-Analytic Exciton Dispersion and Dimensionality effect}\textbf{Exciton
dispersions from the HREELS measurements.} (a), (b) Normalized HREELS
mapping for Nb\protect\textsubscript{3}Cl\protect\textsubscript{8}
at 300 K ($\alpha$ phase) and Nb\protect\textsubscript{3}Cl\protect\textsubscript{2}Br\protect\textsubscript{6}
at 106 K ($\beta$ phase). The white dashed lines represent the fitting
results of the momentum-dependent EDCs. (c), (d) Extracted dispersions
of the inter-band transitions for the $\alpha$ phase and $\beta$
phase, obtained by fitting the experimental data. (e), (f) The second
derivative images of the inter-band transitions for the $\alpha$
phase and $\beta$ phase. (g), (h) Extracted dispersions of the excitons
for the $\alpha$ phase and $\beta$ phase, with the gray curves represent
the linear fit and parabolic fit to the correspond dispersions, respectively.
(i), (j) The second derivative images of the exciton dispersions for
the $\alpha$ phase and $\beta$ phase. Notice, to clearly display
the dispersion shape near the BZ center, the data in (c-j) are symmetrized
along the gamma point to show the negative momentum. The error bars
are determined through the convolution of instrumental resolution
($\sim3.0\ {\rm meV}$) and the standard errors derived from the fitting
process ($\sim0.3\ {\rm meV}$ for the $\alpha$ phase and $\sim5.0\ {\rm meV}$
for the $\beta$ phase).}
\end{figure*}

\emph{Dimensionality-modulated Exciton Dispersion.} To better visualize
the dispersion of the observed excitations, we initially normalized
the HREELS mapping for both the $\alpha$ phase {[}Fig. \ref{fig:Non-Analytic Exciton Dispersion and Dimensionality effect}(a){]}
and the $\beta$ phase {[}Fig. \ref{fig:Non-Analytic Exciton Dispersion and Dimensionality effect}(b){]}
by integrating the intensity over the 0.5--1.5 eV energy range at
each momentum $\boldsymbol{q}$ \footnote{In HREELS measurements, incident electrons interact with the long-range
dipole field generated by electronic excitations within the material
\citep{RN649,RN650}. As a result, the loss signals are confined to
the dipole scattering region, leading to a rapid attenuation of signal
intensity as the scattering angle deviates from specular scattering.
Consequently, this limits visualization of electronic excitation dispersion
in raw HREELS spectra.}. From the normalized HREELS spectra, it is evident that both the
IT and the EX features exhibit clear dispersion within the concerned
momentum range.

To quantitatively capture the dispersions, we employed a Voigt function
to fit the momentum-dependent EDCs. Additionally, to further enhance
the clarity of the dispersion near the BZ center, we applied a second
derivative to the normalized 2D mapping spectra, which helps highlight
the dispersion shape. The fitted dispersion and the second derivative
mapping in both the $\alpha$ phase and $\beta$ phase show consistent
results. First, we investigated the dispersions of the inter-band
transitions in both the $\alpha$ phase {[}Fig. \ref{fig:Non-Analytic Exciton Dispersion and Dimensionality effect}(c)
and (e){]} and $\beta$ phase {[}Fig. \ref{fig:Non-Analytic Exciton Dispersion and Dimensionality effect}(d)
and (f){]}. Within the concerned momentum range ($|\boldsymbol{q}|\sim0.06\ \text{\AA}^{-1}$),
the inter-band transitions both display parabolic-like dispersions
near the BZ center, with a flat energy distribution less than 20 meV.
Since inter-band transitions result from the collective contributions
of bands throughout the entire BZ, these flat dispersions further
consolidate the flatness of the LHB and UHB in both the $\alpha$
and $\beta$ phases.

Next, we focus on the exciton dispersions. In the $\beta$ phase,
the two excitonic bands, EX1 and EX2, both exhibit clear parabolic
dispersion, as evidenced by both the fitted results {[}Fig. \ref{fig:Non-Analytic Exciton Dispersion and Dimensionality effect}(h){]}
and the second derivative results {[}Fig. \ref{fig:Non-Analytic Exciton Dispersion and Dimensionality effect}(j){]}.
These results undoubtedly reproduce the expected dispersions of 3D
Wannier excitons, consistent with the 3D nature of the $\beta$ phase.
In contrast, in the $\alpha$ phase, the exciton displays a distinctive
``V''-shaped dispersion near the $\overline{\Gamma}$ point {[}Fig.
\ref{fig:Non-Analytic Exciton Dispersion and Dimensionality effect}(g)
and (i){]}, indicating that it is a 2D massless exciton. Within the
small momentum range ($|\boldsymbol{q}|\sim0.025\ \text{\AA}^{-1}$),
the exciton exhibits a clear linear dispersion, which gradually deviates
from linearity as $|\boldsymbol{q}|$ increases. A linear fit of the
dispersion near the BZ center ($|\boldsymbol{q}|\sim0.025\ \text{\AA}^{-1}$)
is given by: $E=E_{0}+v\cdot|\boldsymbol{q}|$, where the slope $v=0.51\ {\rm eV}/\lyxmathsym{\AA}$
represents the exciton group velocity, directly reflecting the exciton
diffusion speed \citep{RN221,RN595}.

To fully understand the massless exciton dispersion in the $\alpha$
phase and capture the dramatic changes in dispersion across the phase
transition, we employ a modeled Hamiltonian that accounts for the
quantum effects of exchange scattering between electron-hole pairs
\citep{RN221,RN238,RN543}. In this framework, the exciton dispersion
is obtained by solving the Bethe Salpeter equation in the electron-hole
basis \citep{RN655,RN656,RN657}. The solution to the Hamiltonian
results from the combined effect of the electron-hole transition energy
($E_{k}$), the direct interaction ($E_{d}$), and the exchange interaction
($E_{ex}$). Here, $E_{k}$ contributes to the exciton kinetic energy,
while $E_{d}$ and $E_{ex}$ contribute to the exciton potential energy.
Using $\boldsymbol{q}\cdot\boldsymbol{p}$ perturbation theory with
the modeled Hamiltonian, the contributions of these three interactions
in the long-wavelength limit ($|\boldsymbol{q}|\rightarrow0$) can
be expressed as:
\begin{equation}
E_{k}(\boldsymbol{q})\propto|\boldsymbol{q}|^{2},
\end{equation}
\begin{equation}
E_{d}(\boldsymbol{q})\propto-|\boldsymbol{q}|^{2},
\end{equation}
\begin{equation}
E_{ex}(\boldsymbol{q})\propto\begin{cases}
\cos^{2}(\theta_{q})+\chi|\boldsymbol{q}|^{2}, & \text{\text{(in 3D)}}\\
|\boldsymbol{q}|\cdot\cos^{2}(\theta_{q})+\kappa|\boldsymbol{q}|^{2}. & \text{(in 2D)}
\end{cases}
\end{equation}
where $\theta_{q}$ is the angle between the momentum $\boldsymbol{q}$
and the dipole matrix element $\boldsymbol{P}_{S}$ of the exciton
state $|S(\boldsymbol{q})\rangle$, and the factors $\chi$, $\kappa$
are constants related to the specific band structure of the system.
The exciton dispersion near the $\overline{\Gamma}$ point is determined
by the competition between these three terms. In 3D systems, the exciton
dispersion always manifests a characteristic parabolic shape. In contrast,
for 2D systems, the presence of the linear term in $E_{ex}$ induces
a distinct linear exciton dispersion near the BZ center, as illustrated
in Fig. \ref{fig:Schematic of Dimensionality effect}(d). When the
parabolic contributions arising from $E_{k}$ and $E_{d}$ are dominated
by a strong $E_{ex}$ term, the linear behavior of the exciton dispersion
persists over an extended range from the BZ center. This enhanced
linear regime becomes more accessible for experimental observation
through HREELS.

Nb\textsubscript{3}Cl\textsubscript{8} is a Mott insulator with
relatively flat LHB and UHB. As a result, the kinetic term $E_{k}$
can be treated as a constant, independent of the momentum $\boldsymbol{q}$,
which is consistent with the observed flat dispersion of the IT {[}Fig.
\ref{fig:Non-Analytic Exciton Dispersion and Dimensionality effect}(c-f){]}.
The breathing Kagome lattice structure, which promotes significant
overlap of the electron-hole wavefunctions, further suppresses the
Coulomb screening \citep{RN639,RN634}. Consequently, Nb\textsubscript{3}Cl\textsubscript{8}
exhibits enhanced exchange interactions, i.e., $E_{ex}$ dominates
the exciton dynamics, providing the primary mechanism behind the observed
dimensional dependence of the exciton dispersions. The spontaneous
structural phase transition, which notably alters the interlayer coupling
strength, establishes Nb\textsubscript{3}Cl\textsubscript{8} as
an ideal natural platform for investigating excitonic dimensional
effects. In the $\alpha$ phase, the strong vdW nature of Nb\textsubscript{3}Cl\textsubscript{8}
results in negligible interlayer coupling, and thus excitons are confined
within the 2D layers, with the in-plane screening effect \citep{RN608,RN589,RN580}
predominantly contributing to the 2D massless linear exciton dispersion.
In contrast, in the $\beta$ phase, the strong interlayer hybridization
alters the electronic properties, causing the excitons to no longer
be localized within 2D layers. This leads to a more pronounced out-of-plane
screening effect, resulting in exciton dispersion that exhibits 3D
characteristics.

In summary, we observed exciton splitting across the phase transition
of Nb\textsubscript{3}Cl\textsubscript{8}, originating from bonding-antibonding
splitting due to enhanced interlayer coupling. More importantly, in
the $\alpha$ phase, we definitively demonstrate the presence of a
quasi-2D massless exciton dispersion, which transitions to a 3D parabolic
dispersion due to the significant enhancement of interlayer coupling
across the phase transition. This unique variation in interlayer coupling
strength makes Nb\textsubscript{3}Cl\textsubscript{8} an ideal platform
for studying the dimensional effects on excitonic band structures.
Beyond the qualitative similarities to 2D semiconductors, the correlated
nature of Nb\textsubscript{3}Cl\textsubscript{8} necessitates advanced
many-body treatments. We expect these results to catalyze future theoretical
and experimental efforts to clarify the unique excitonic mechanisms
in the correlated regime.
\begin{acknowledgments}
This work was supported by the National Key R\&D Program of China
(Nos. 2021YFA1400200, 2022YFA1403000, 2022YFA1403800, and 2023YFA1406500),
the National Natural Science Foundation of China (Nos. 12274446, 12174334
and 12274459), and the Strategic Priority Research Program of the
Chinese Academy of Sciences (No. XDB33000000).
\end{acknowledgments}

\bibliographystyle{apsrev4-2}
\bibliography{4E__IOP_Research_manuscripts_Nb3Cl8_PRL_Final_Submission_Exciton}

\section*{End Matter}

\begin{figure*}
\includegraphics[width=0.95\textwidth]{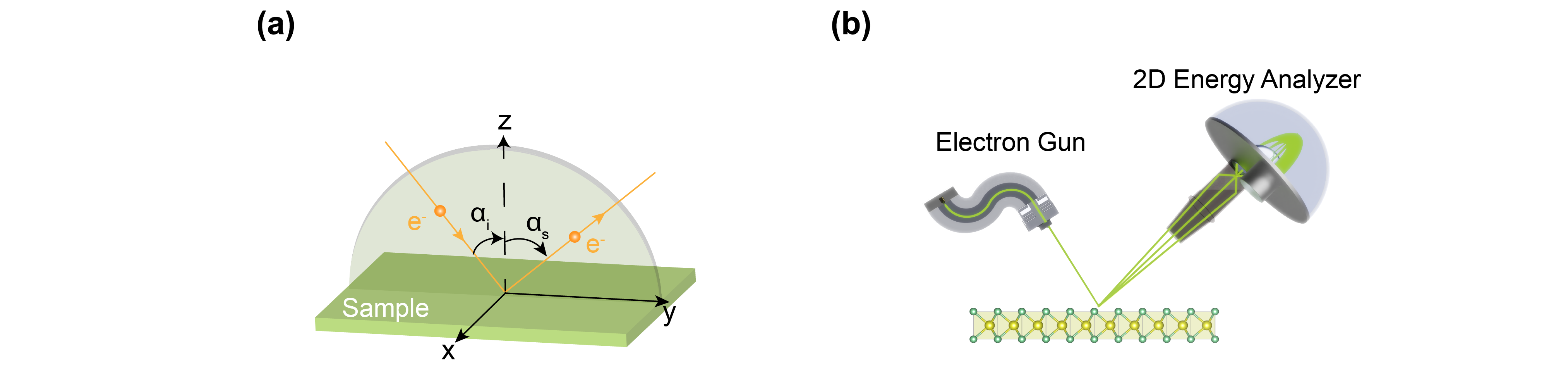}

\caption{\label{fig:Instrument}(a) Schematic of the scattering geometry. (b)
Schematic of the experimental setup of the 2D HREELS.}
\end{figure*}
\emph{Principle of the 2D HREELS.} HREELS probes elementary excitations
via the inelastic scattering of a monochromatic electron beam \citep{raether2006excitation}.
In traditional HREELS measurements, which employ a fixed scattering
geometry (with fixed incident and scattering angles), an EDC is acquired
at a single momentum point. By analyzing the energy and angle of the
monochromatic electron beam before and after scattering, the energy
of the elementary excitation and the momentum parallel to the sample
surface can be determined based on the principles of energy and momentum
conservation {[}Fig. \ref{fig:Instrument}(a){]} \citep{RN650}:

\begin{equation}
E_{Loss}=E_{i}-E_{s},
\end{equation}
\begin{equation}
q_{\parallel}=\frac{\sqrt{2mE_{i}}}{\hbar}\left(\sin\alpha_{i}-\sqrt{1-\frac{E_{Loss}}{E_{i}}}\sin\alpha_{s}\right).
\end{equation}

\noindent where $E_{i}$ and $E_{s}$ are the energy of incident and
scattering electrons, in sequence. To acquire the dispersion of elementary
excitations, it is necessary to change the momentum points by rotating
the angles of the analyzer, sample, and monochromator.

In contrast, our HREELS setup is equipped with a hemispherical energy
analyzer, enabling 2D mapping of momentum and energy simultaneously
without the need to rotate the sample or analyzer, significantly enhancing
data acquisition efficiency {[}Fig. \ref{fig:Instrument}(b){]}. Under
this instrumental configuration, the ultimate energy and momentum
resolutions can reach $0.7\ {\rm meV}$ and $0.002\ \text{\AA}^{-1}$,
respectively \citep{RN604}. In the present work, to optimize detection
efficiency and ensure a high signal-to-noise ratio for the observed
excitonic features, we utilized an operating energy resolution of
$3.0\ {\rm meV}$ and a momentum resolution of $0.005\ \text{\AA}^{-1}$.
\textit{In-situ} HREELS measurements in this study were performed
using a monochromatic electron beam with an incident angle of 60�
and an incident energy of 110 eV.

\emph{Probing Depth of HREELS.} While HREELS is typically characterized
as a surface-sensitive technique, the effective probing depth depends
fundamentally on the underlying scattering mechanism. It is essential
to distinguish between impact scattering and dipole scattering. In
the impact scattering regime, the probing depth is governed by the
inelastic mean free path of the probe electrons---typically $5-10\ \lyxmathsym{\AA}$
for incident energies in the $50-100\ {\rm eV}$ range---rendering
the technique strictly sensitive to the topmost atomic layers.

In contrast, the dipole scattering mechanism, which dominates the
excitonic measurements near the $\Gamma$ point in this study, involves
long-range electromagnetic interactions. In this regime, the probing
depth is independent of the inelastic mean free path. The HREELS cross-section
is directly proportional to the spectral function $S(\boldsymbol{q},\omega)$,
which describes the collective electronic response of the system \citep{RN12}:

\[
S(\boldsymbol{q},\omega)=\int_{-\infty}^{0}S(\boldsymbol{q},z,z',\omega)e^{-|\boldsymbol{q}|\cdot|z+z'|}dzdz'
\]

\noindent where $z$ and $z'$ are coordinates perpendicular to the
surface. The term $S(\boldsymbol{q},z,z',\omega)$ represents the
density-density correlation function:

\begin{gather*}
S(\boldsymbol{q},z,z',\omega)=\sum_{m,n}\left[\langle m|\hat{\rho}(\boldsymbol{q},\boldsymbol{z})|n\rangle\cdot\langle n|\hat{\rho}(-\boldsymbol{q},z')|m\rangle P_{m}\right]\\
\cdot\delta(E-E_{n}+E_{m})
\end{gather*}

\noindent where $\hat{\rho}$ is the charge density operator, $P_{m}=e^{-E_{m}/k_{B}T}/Z$
is the Boltzmann factor, and $|m\rangle$, $|n\rangle$ denote the
many-body states of the sample with energies $E_{m}$ and $E_{n}$,
respectively.

Crucially, the weighting factor $e^{-|\boldsymbol{q}|\cdot|z+z'|}$
in the integral dictates that the effective probing depth scales as
$\sim1/|\boldsymbol{q}|$. Within our investigated momentum range
($0.00$ to $0.06\ \text{\AA}^{-1}$), the probing depth extends significantly
beneath the surface, thereby encompassing multiple van der Waals layers.
Consequently, the excitonic signals detected in Nb\textsubscript{3}Cl\textsubscript{8}
originate from the collective response of many near-surface layers,
effectively representing the intrinsic bulk-like properties of the
material. This ensures that the observed evolution of exciton dispersion
is a manifestation of dimensionality-driven effects rather than a
localized surface phenomenon.
\end{document}